# Is there a neutrino speed anomaly?

J. Knobloch[1]

**Abstract**

The OPERA collaboration reported [1] an anomaly in the ratio of the speed of neutrinos to the speed of light of (v-c)/c =(2.48±0.28 (stat.) ± 0.30 (sys.))×$10^{-5}$.
I identify sources of systematic uncertainty that were not considered in the report.
A basic assumption in [1] is that the proton time structure represents the time structure of the neutrino flux. In this manuscript, I argue that this assumption can be questioned on the basis of the information available. These additional uncertainties can invalidate the claimed neutrino speed anomaly.

**Introduction**

In a speed measurement there are obviously two major elements: distance and time. Both have been covered by the OPERA collaboration in [1] and, not being an expert on geodesy or GPS time measurement, I refrain from commenting on these. In the OPERA analysis there is, however, a third element that did not give rise to detailed consideration in [1]: The measurement of the time structure or Particle Density Function (PDF) of the neutrinos emanating from the CERN CNGS (CERN Neutrinos to Gran Sasso) system. The proton extraction lasts for about 10.5 µs while the claimed time leading to the anomaly is 60.7 ns measured with a reported accuracy of ±6.9 ns (stat.) and ±7.4 ns (sys.). Particularly important to the measurement are the leading and trailing edges of the neutrino time distribution. The OPERA analysis assumes that the proton PDF is measured correctly and that it represents exactly the neutrino PDF. In the following, I argue that both assumptions can be questioned and that systematic effects of the order of the observed anomaly have been neglected.

**Correct measurement of the proton PDF**

The proton PDF is measured by a beam current transformer (BCT), a coil coaxial with the beam where the passing protons induce a current proportional to the proton flux. The signal of the BCT is digitized by a digitizer operating at 1 giga-samples/s. In a thesis [2], additional details of the analysis are given. For certain run periods, the digitizer did not perform correctly by either saturating the signal or by inducing oscillations. These periods have been removed from the analysis. It should however be noted that an oscillating 30 and 60 ns structure is observed in the final waveforms, most pronounced during the last quarter of the extractions and in particular over the falling edge of the proton spill, see Fig. 8.4 of [2]. Such oscillations are still visible after summing several hundred individual measurements. These oscillations significantly deform part of the proton PDF and can therefore give rise to additional systematic errors.
In the analysis described in [2], the oscillations are "eliminated" – one should rather say attenuated - by a low-band software filter of 8 MHz. Such a filter not only attenuates the noise but also influences the leading and trailing edges of the proton PDF, which are instrumental to the final result.
The proton PDF has a leading edge rise time of about 800 ns and a trailing edge fall time of about 400 ns (see Fig. 12 of [1]) and a more or less flat top in between:

---

[1] The author is retired staff member from CERN. The opinions expressed in this article are the responsibility of the author alone and do not incur any liability to CERN.

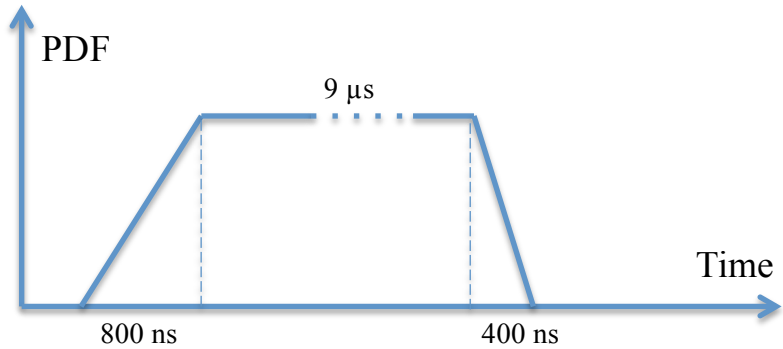

The low-band filter of 8 MHz will distort the leading and trailing edges with a time constant of 1/8,000,000 = 125 ns as shown schematically here:

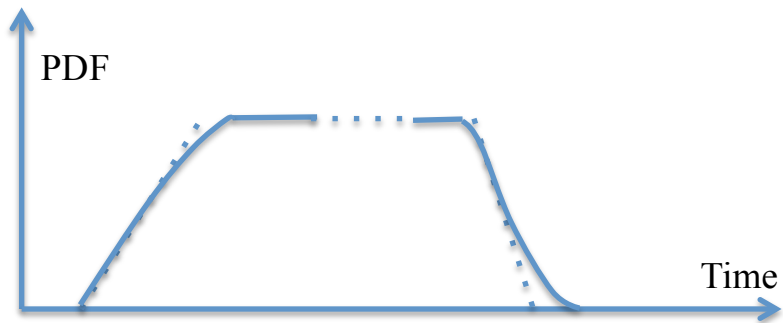

The precise shape will depend on the filtering algorithm used. A deformation of the PDF edges by 125 ns will certainly have an impact on the total anomaly of 60 ns.

**Broadening of the PDF**

The proton PDF used for the analysis is a sum of the individual BCT measurements that correspond to neutrino events in OPERA. For the time alignment of the individual distributions, the trigger signal of the kicker magnet MKE4 is used. The timing of this trigger (kick delay) is optimized (in steps of 100 ns) in order to minimize beam loss, in particular in the septum magnet. It may happen that after such optimization or after a machine development period, this delay does not come back to the previous value. If this would happen during the yearly data taking, some fraction of the proton distributions would be shifted by e.g. 100 ns. This effect will lead to a broadening of the summed PDF.
The distribution of events in OPERA is not affected by this effect because detailed time corrections are applied to the neutrino events individually, in particular a correction for the GPS time recorded with each proton waveform adjusted for the daily excursion of the GPS clocks of 60 ns as shown in Fig. 9.8 of [2].
It can therefore not be excluded that the width of the used proton PDF is larger than that of the neutrino event distribution.
One may argue that this broadening would not change the mean of the distribution and would therefore not impact the result.
As mentioned above, however, the leading and trailing slopes differ by about a factor 2. As the steeper slope will have a larger impact on the fit result, this will lead to a shift in the final measurement. Visually, Fig. 12 of [1] does not allow to exclude a broadening of the PDF by the order of 40 ns.

**Differences between proton and neutrino PDF**

Assuming that the proton PDF is correctly measured, there are mechanisms that alter the shape of the neutrino PDF. If the kicker magnet strength is not constant during the spill, the proton beam will not impact the target in a single fixed point but move about the surface of the target. Another example, a variation of the strength of the focussing elements (Horn and Reflector) over the spill time, was elaborated in previous versions of this document. Because the impact is probably negligible, it is no longer included here.

**Conclusion and outlook**

It should be pointed out that the arguments presented here are based on available publications. It may well be that some of my points would not withstand the scrutiny of the OPERA collaboration. I do believe, though, that effects of the PDF shape should be entered into the list of systematic uncertainties and may even constitute the single largest contribution.

The impact of the 30 and 60 ns oscillations and of the low-band filter can probably be evaluated by introducing them into a full simulation of the analysis.
The origin of these oscillations needs to be understood at the level of the BCT hardware and the digitization electronics providing a better assessment of the impact on the measurement.
The PDF broadening could be estimated by an additional fit parameter or by fitting the two slopes separately.
A final conclusion on the shape of the neutrino PDF would be a measurement of the time structure of the muon flux after the decay region.

I conclude that a possible difference between the proton and neutrino PDF was not sufficiently considered in evaluating the systematic uncertainties summarized in Table 2 of [1]. A more detailed analysis may lead to systematic errors impacting the significance of the result.